\documentclass[12pt]{article}
\usepackage{amssymb} 
\parskip=\medskipamount

\def\a{\alpha}
\def\b{\beta}

\def\d{\delta}
\def\e{\epsilon}
\def\g{\gamma}

\def\q{\theta}

\def\C#1{\overline{#1}}

\def\dt#1{{\buildrel{\hbox{\LARGE .}} \over {#1}}}   
\def\da{{\dt\alpha}} 
\def\db{{\dt\beta}}
\def\dg{{\dt\gamma}}
\def\dd{{\dt\delta}}

\def\cd{{\cal D}}

\def\bC{{\mathbb C}}

\def\bZ{{\mathbb Z}}

\newfont{\goth}{eufm10 scaled \magstep1}

\def\gg{\mbox{\goth g}}
\def\gh{\mbox{\goth h}}

\def\SO#1{{\mathrm{SO(#1)}}}

\def\n{\nabla}
\def\op{\oplus} 

\def\ra{\rightarrow}

\def\fr#1#2{{\textstyle{#1\over #2}}} 
\def\be{\begin{equation}}
\def\ee{\end{equation}}
\def\re#1{(\ref{#1})}
\def\la#1{\label{#1}}  
\def\arr{\begin{array}{rll}}
\def\ea{\end{array}}
\def\bea{\begin{eqnarray}}
\def\eea{\end{eqnarray}}

\def\square{\kern1pt\vbox
            {\hrule height 0.6pt\hbox{\vrule width 0.6pt\hskip 3pt
 \vbox{\vskip 6pt}\hskip 3pt\vrule width 0.6pt}\hrule height 0.6pt}\kern1pt}

\def\sp{super Poincar\'e algebra}
\def\sds{super de Sitter algebra}

\font\cmsslll=cmss10 at 14 pt

\begin{document}
\begin{titlepage}
\rightline{hep-th/0010056}
\vskip 5 cm
\begin{center}
{\huge   Democratic Supersymmetry}
\vskip 1 true cm
{\cmsslll Chandrashekar Devchand$^a$ and Jean Nuyts$^b$}
\vskip 0.8 true cm
{\small  devchand@math.uni-bonn.de\ ,\ Jean.Nuyts@umh.ac.be
\vskip 0.2 true cm
{\it $^a$ Mathematisches Institut der Universit\"at Bonn,
Beringstra\ss e 1, D-53115 Bonn, Germany}\\[2pt]
{\it $^b$  Physique Th\'eorique et Math\'ematique, 
Universit\'e de Mons-Hainaut}\\[1pt]
{\it 20 Place du Parc, B-7000 Mons, Belgium}}
\end{center}
\vskip 3 true cm
\begin{quote}
{\bf Abstract\,:} We present generalisations of N-extended supersymmetry 
algebras in four dimensions, using Lorentz covariance and invariance under 
permutation of the N supercharges as selection criteria. 
\end{quote}
\end{titlepage}
\vfill\newpage
\def\thesection{\Roman{section}}

\section{Introduction}

Recent developments in string theory have revealed the need to study
generalisations of supersymmetry which lie beyond the realm of existing
classifications of spacetime supersymmetry algebras. 
Spacetime supersymmetry algebras are $\bZ_2$-graded super Lie algebras 
$\ \gg = \gg_0 \op \gg_1\ $ having even $\,\gg_0\,$  and odd $\,\gg_1\,$ 
subspaces, where the even part 
$\ \gg_0 =\ {<}M{>}\ \op\ {<}P{>}\ \op\ \gh\ $ includes 
the generators of spacetime Lorentz transformations $M$, translations $P$ 
and a subspace of additional `internal symmetries' $\gh$. 
The usual relation between spin and statistics
implies that generators of $\gg_1$ transform as half-integer spin 
representations under the Lorentz transformations.
Traditional classifications of spacetime supersymmetries were based
on assumptions arising from the additional requirement that the
supersymmetries act on either S-matrix elements \cite{HLS} or on some
physical Hilbert space of particle states \cite{N}. 
In particular these restrict the maximum spin of the generators to be one
and require the internal symmetries to be `central' in the sense that
they commute with all other generators. Moreover, in four dimensional
space-time, the realisation
of these algebras on physical states restricts finite dimensional 
representations to contain fields of spin less than or equal to two 
and the maximal number $N$ of independent supercharges in $\gg_1$ to eight. 

There are several instances in which spacetime supersymmetries and
representations more general than those allowed in traditional settings 
occur. In M-theory, for instance, the internal symmetries $\gh$ do not commute 
with the Lorentz generators (see e.g. \cite{H}). 
In $N$-extended super self-dual theories in 
four dimensional Euclidean space, 
finite dimensional representations containing fields of spin higher than two 
do occur and there are consistent theories for any choice of $N$ \cite{DO}.
In $N{=}2$ string theory \cite{DL}, the absence of the usual relation
between spin and statistics gives rise to a realisation of a purely even
variant of supersymmetry \cite{AC} on an infinite dimensional space of
string states. There are indications that this statistics-twisted version
of supersymmetry is related to an $N\ra\infty$ extension of the \sp, 
which has a realisation
on an $N{=}\infty$ self-dual Yang-Mills supermultiplet \cite{DO}. 
These examples show that there seems to be room for the study of more 
general superalgebras
containing the ($N$-extended) \sp\ or the \sds\ as a subalgebra or as 
a contraction.  The work of Fradkin and Vasiliev (e.g. \cite{FV}) on higher
spin superalgebras on anti de Sitter space is also noteworthy in this respect. The present paper is a further contribution in this direction.

In a series of recent papers \cite{DN} we recently
developed an approach to the study of generalised super-Poincar\'e algebras
containing generators having spins higher than one. We showed that, 
contrary to common belief, such superalgebras indeed exist and are
realisable in terms of vector fields on generalised superspaces having 
coordinates of higher spin which commute or anticommute in accordance 
with their statistics.  We constructed 
numerous examples of generalised superalgebras with generators having spins 
up to two. 

In the present paper, we address ourselves to another type of generalisation,
concerning the question of higher multiplicities of certain representations
in the superalgebra. Theories with $N$ supercharges
are of special interest. In these, there does not seem to be any principle
which distinguishes some of the supercharges from the others and field 
theories containing such supercharges are usually taken to be invariant
under permutation of the supercharges. In this paper, we impose this 
permutation invariance at the level of the superalgebra, 
introducing what we will call {\it democratic superalgebras}.
Our purpose here is not a complete classification of possibilities;
rather, we aim to show that under the imposition of {\it democracy}, 
even in the widely familiar four-dimensional case,
an investigation of super Jacobi identities yields some potentially
interesting democratic spacetime superalgebras which lie beyond 
known classifications.
The main novel feature which arises in our approach is that the algebra of 
Lorentz scalars $\gh$ generated by the superderivations is no longer either 
Abelian or in the centre of $\gg$. Although democracy implies the 
Coleman-Mandula requirement \cite{CM} that the scalars commute with 
(even) translations, they possibly rotate spinor derivations among themselves.

\section{Democratic Superalgebras}

\subsection{Four dimensional spacetime supersymmetry}

Since our aim is to generalise traditional discussions and 
since our considerations are purely algebraic, we restrict ourselves to the
general complex setting.
The question of the appropriate real form depends in any case on the 
signature of the space-time on which the superalgebra is to be realised; 
and this depends on the specific context of the application.
We consider the Lorentz group to be $\SO{4,\bC}$,
with complex generators $ M_{\a\b}, M_{\da\db}$, where there is of course
no conjugation between dotted and undotted spinor indices.

We shall consider  $\bZ_2$-graded $N$-extended complex supersymmetry 
algebras of the form $\gg = \gg_0 \op \gg_1$,
with even part 
\be
\gg_0\  =\ \  < M_{\a\b}, M_{\da\db} , \n_{\a\da} >\ \op\ \gh\ ,
\ee
where $\n_{\a\da}$ denotes the
derivative vector fields generating translations, and $\gh$ is the
subspace of internal symmetries 
\be
\gh\  =\ \ < Y^i, Z^{ij}=-Z^{ji}\ ;\ 
     \sum_i Y^i = 0,\ \sum_i Z^{ij}=0,\ i,j=1,\ldots N >\ ,
\ee 
spanned by a set of Lorentz scalar generators, $(N{-}1)$ $Y$'s
and $(N{-}1)(N{-}2)/2$ $Z$'s.
 
The odd subspace $\gg_1$ is spanned by $N$ copies of the two types of 
spinor representations of $so(4,\bC)$, namely the $2N$ fermionic 
operators $\n^i_{\a},\n^i_{\da}$ ($i=1,\ldots,N$), which together with the
bosonic vectorial operator $\n_{\a\db}$, form the set of superderivations
acting on an $N$-extended superspace. 
We denote the vector space of superderivations 
$$
\cd =\ < \n_{\a\db}, \n^i_{\a},\n^i_{\da}>\  =\  \cd_0 \op \cd_1\ ,
$$
where the even and odd parts are spanned by the vector and spinor derivations
respectively. The vector space $\cd$ may be extended to include vector fields
having higher spins on the lines of the consideration in \cite{DN}. 
For simplicity, however,  we restrict ourselves here, to the consideration 
of operators having spin less than or equal to one.  

We shall assume that all the elements in $\gg$  have commutation or
anticommutation relations in agreement with their statistics and with 
covariance under the Lorentz 
transformations with generators $M_{\a\b},M_{\da\db},\ \a,\b,\da,\db=1,2$
satisfying
\be\arr
\Bigl[M_{\a\b},M_{\g\d}\Bigr]
          &=& \e_{\b\g}M_{\a\d}+\e_{\a\g}M_{\b\d}
           +\e_{\b\d}M_{\a\g}+\e_{\a\d}M_{\b\g} \ , \\[8pt]
\left[M_{\da\db},M_{\dg\dd}\right]
          &=&\e_{\db\dg}M_{\da\dd}+\e_{\da\dg}M_{\db\dd}
           +\e_{\db\dd}M_{\da\dg}+\e_{\da\dd}M_{\db\dg} \ , \\[8pt]
\left[M_{\a\b},M_{\dg\dd}\right]
          &=&0 \ \ .
\label{comMM}
\ea\ee
Lorentz covariance, in particular, determines all commutators of the 
basic operators with the $M$, namely 
\bea   
\Bigl[M_{\a\b},\n^i_{\g}\Bigr]
                  =\e_{\a\g}\n^i_{\b}+\e_{\b\g}\n^i_{\a} \ ,
      \ \  \left[M_{\a\b},\n^i_{\dg}\right]
                  =0
          &,& 
      \ \  \Bigl[M_{\a\b},Y_i\Bigr]=0\ ,
      \ \  \Bigl[M_{\a\b},Z^{ij}\Bigr]=0
   \nonumber\\
\left[M_{\da\db},\n^i_{\dg}\right]
                  =\e_{\da\dg}\n^i_{\db}+\e_{\db\dg}\n^i_{\da}\ ,
      \ \   \left[M_{\da\db},\n^i_{\g}\right]
                  =0 
           &,& 
      \ \  \left[M_{\da\db},Y_i\right]=0\ ,
      \ \  \left[M_{\da\db},Z^{ij}\right]=0\ .
\label{comMop}
\eea
Given these commutation rules, all Jacobi identities involving at least 
two $M$'s are automatically satisfied. Lorentz covariance also yields
restrictions on the (anti)commutators of any two elements of $\gg$. 
These guarantee that the Jacobi identities involving at least one $M$ are 
also automatically satisfied.  

The spinor derivations $\n^i_{\a},\n^i_{\da}$ are taken to transform
under some group of automorphisms $T$ of the superalgebra $\gg$:
\be
       T\n^i_{\a}T^{-1}=U^i_{\phantom{i}j}\,\n^j_{\a}\ ,
\ \    T\n^i_{\da}T^{-1}=V^i_{\phantom{i}j}\,\n^j_{\da}\ ,
\label{aut}
\ee
where the matrices $U,V$ are representations of the group element $T$.
In this paper, we make particular use of discrete transformations, 
taking $U$ and $V$ to be permutation matrices on the index $i$. 
When the automorphism
group is continuous, the action of the group can be expressed in the form of 
commutation relations with the generators of the group: for instance the 
scalar generators $Y$ or $Z$ which appear in \re{YS}-\re{ZD}.

We shall also allow the possibility of generating scalars by anticommuting 
spinor derivations, e.g. 
$\left\{ \n^i_\a , \n^j_\b \right\} \sim \e_{\a\b} Z^{ij}$.
Traditionally \cite{HLS}, such Lorentz scalars 
are always taken to be central with respect to $\cd$.
In our approach we do not a priori restrict the Lorentz scalars to
be central. In fact they rotate the spinor derivations 
$\n^i_{\a},\n^i_{\da}$ just as the automorphisms \re{aut}. This is
the main source of our novel examples of spacetime supersymmetries. 

\subsection{Democracy}

\subsubsection{Permutation invariants}

We shall impose what we call democracy: we require the
supercommutation relations to be invariant under the
combined permutations of the $i$-indices of $\n^i_{\a}$ and of
$\n^i_{\da}$. The group generating democracy $S_N$ is the diagonal group of 
two groups of permutations
acting independently on the two sets of spinors, with permutation matrices 
$U=V$ in \re{aut}.

The Clebsch-Gordon coefficients of the democratic group may be described 
as follows. The permutation invariant coupling among $p$ ($ p>1$)
$i$ type indices can be associated to Young-type diagrams. 
Given a Young diagram with $p$ ($p>0$)
boxes denoted $[m]=[m_1,m_2,\ldots,m_p]$, with $m_j$ boxes in the
$j$-th row ($\sum_j m_j=p,\ m_{i+1}\leq m_i$), we associate to it a
$p$-index tensor $\q_{[m_1 m_2\ldots m_p]}^{i_1 i_2\ldots i_p}$ defined by 
\bea
\q_{[m_1 m_2 \ldots m_p]}^{\overbrace{i_1 \ldots i_{m_1}}^{m_1} 
 \overbrace{j_{1}\ldots j_{m_2}}^{m_2}
 \ \overbrace{k_{1}\ldots k_{m_3}}^{m_3} 
 \ \ldots}
   &=& 1\quad{\rm{if}}\ i_{1}{=}\ldots{=}i_{m_1},\ j_{1}{=}\ldots{=}j_{m_2},
                                        \ k_{1}{=}\ldots{=}k_{m_3},\ldots
     \nonumber\\
   &=&0\quad {\rm{otherwise}}\ \ .
\eea
Note that these tensors clearly do not have the standard Young diagram symmetries.
{}From these $\q$ tensors, by permuting indices, all the invariant tensors
of the permutation group can be constructed. For a Young-type diagram with
$p$ boxes,
if $n_l$ is the number of rows having length $m_k=l$,
the number of independent invariant tensors is given by
$p!/(\prod_k{m_k!}\prod_l{n_l!})$.

Some of these tensors have a simple interpretation in terms
of the familiar Kronecker tensor $\d^{ij}$. In particular
\bea
\q_{[2]}^{ij}&=& \d^{ij}
     \nonumber\\
\q_{[22]}^{ijkl}&=& \d^{ij}\d^{kl}
     \nonumber\\
\q_{[222]}^{ijklmn}&=& \d^{ij}\d^{kl}\d^{mn}
\eea
correspond to the invariant tensors of $so(N)$. 
These are special cases of the identities
\be
\q_{[m_{1} m_{2} m_{3} \ldots]}^{i_{1}
\ldots i_{m_{1}}i_{m_1+1}\ldots i_{m_1+m_2}\ldots} 
 =    \q_{[m_1]}^{i_1\ldots i_{m_1}}
 \ \q_{[m_2]}^{i_{m_1+1}\ldots i_{m_1+m_2}}\ \q_{[m3]}^{\ldots}\ \ldots
\ \ .
\ee
One further useful identity (with summation over repeated indices assumed) is
\be
\q^{i}_{[1]}\q^{ijk\ldots}_{[m]}= \q^{jk\ldots}_{[m-1]}\ \ ,
\ee
which is valid for all $m>0$ if we define
\be
\q_{[0]}:=N\ \ .
\ee

\subsubsection{Trace conditions}

We note that the tensor $\q^{i}_{[1]}$ can be used to decompose tensors 
into their permutation irreducible parts. In particular, a vector $V^{i}$ 
has two irreducible components given by the scalar projection $S$
\be 
S=\q^{j}_{[1]} V^{j}
\la{irscalar}
\ee
and its complementary piece, of dimension $N{-}1$,
\be
Y^{j}=V^{i}-\fr{1}{N}\q^{i}_{[1]} S\ \ . 
\la{irvector}
\ee
Similarly, a general antisymmetric tensor $T^{ij}$ can be decomposed under 
the permutation group into two irreducible pieces. A piece of the form $Y^{j}$ 
is obtained by the projection 
\be
Y^{j}=\q^{i}_{[1]} T^{ij}\ ,
\la{irY}
\ee
which satisfies
\be
\q^{j}_{[1]} Y^{j}=0 \ \ . 
\la{irYeq}
\ee
The other irreducible piece $Z^{ij}$, of dimension
$(N{-}1)(N{-}2)/2$, can be defined by 
\be
Z^{ij}=T^{ij}-\fr{1}{N}\left(\q^{i}_{[1]}Y^{j}-\q^{j}_{[1]}Y^{i}\right)
\la{irZ}
\ee
and satisfies
\be
\q^{i}_{[1]} Z^{ij}=0 \ \ .
\la{irZeq}
\ee
We will generically call conditions imposed on the 
structure constants which guarantee the irreducibility of the relevant tensors
{\it trace conditions}. The tensors $Y$ in \re{irvector} and $Z$ in 
\re{irZ} will be called {\it{trace-free}}.

Since the $Y^{k}$ and the $Z^{kl}$ 
need to satisfy the trace-free conditions (\ref{irZeq}) and (\ref{irYeq}), 
it is convenient to use some partially 
trace-free combinations of the invariant 
$\q$-tensors with certain symmetries
\bea
t^{ij}_{[2]} &:=&  \q^{ij}_{[2]} - \fr{1}{N}  \q^{ij}_{[11]} 
      \nonumber \\[8pt]
t^{ijk}_{[21]} &:=&  \q^{ijk}_{[21]} - \fr{1}{N}  \q^{ijk}_{[111]} 
      \nonumber \\[8pt]
t^{ijk}_{[3]} &:=&  \q^{ijk}_{[3]} - \fr{1}{N} 
        \left( \q^{kji}_{[21]}+\q^{ijk}_{[21]}+\q^{ikj}_{[21]}\right) 
      \nonumber \\[8pt]
t^{ijkl}_{[22]} &:=&  \q^{ijkl}_{[22]} -  \q^{kjil}_{[22]} 
           - \fr{1}{N} \left(  \q^{ijkl}_{[211]} - \q^{kjil}_{[211]}
                             - \q^{ilkj}_{[211]} + \q^{klij}_{[211]} \right)
      \nonumber \\[8pt]
t^{ijklm}_{[32]} &:=&\q_{[32]}^{jklim}-\q_{[32]}^{ikljm}
                       -\q_{[32]}^{jkmil}+\q_{[32]}^{ikmjl}
       -\fr{1}{N}\left(\q_{[311]}^{jklim}-\q_{[311]}^{ikljm}
                     -\q_{[311]}^{jkmil}+\q_{[311]}^{ikmjl}\right)
       \nonumber \\[6pt]
       &&
       +\fr{1}{N}\left(\q_{[221]}^{jmilk}{-}\q_{[221]}^{imjlk}
       {+}
       \q_{[221]}^{jkilm}{-}\q_{[221]}^{ikjlm}
                      {-}\q_{[221]}^{jkiml}{+}\q_{[221]}^{ikjml}
       {-}\q_{[221]}^{lkimj}{+}\q_{[221]}^{lkjmi}
                      {+}\q_{[221]}^{mkilj}{-}\q_{[221]}^{mkjli}\right)
         \nonumber \\[6pt]
       &&
       -\fr{3}{N^2}\left(\q_{[2111]}^{jmkil}-\q_{[2111]}^{imkjl}
                       -\q_{[2111]}^{jlkim}+\q_{[2111]}^{ilkjm}\right)
         \nonumber \\[8pt]
t^{jkmiln}_{[222]} &:=&    \q_{[222]}^{jkmiln}-\q_{[222]}^{ikmjln}
                          -\q_{[222]}^{jlmikn}+\q_{[222]}^{ilmjkn}
                          -\q_{[222]}^{jknilm}+\q_{[222]}^{iknjlm}
                          +\q_{[222]}^{jlnikm}-\q_{[222]}^{ilnjkm}
      \nonumber \\[6pt]
      &&-\fr{1}{N} \left(    \q_{[2211]}^{jkmiln}{-}\q_{[2211]}^{ikmjln}
                          {-}\q_{[2211]}^{jlmikn}{+}\q_{[2211]}^{ilmjkn}
                          {-}\q_{[2211]}^{jknilm}{+}\q_{[2211]}^{iknjlm}
                          {+}\q_{[2211]}^{jlnikm}{-}\q_{[2211]}^{ilnjkm}\right. 
      \nonumber \\[6pt]
     &&\qquad          {-}\q_{[2211]}^{limkjn}{+}\q_{[2211]}^{kimljn}
                          {+}\q_{[2211]}^{ljmkin}{-}\q_{[2211]}^{kjmlin}
                          {+}\q_{[2211]}^{linkjm}{-}\q_{[2211]}^{kinljm}
                          {-}\q_{[2211]}^{ljnkim}{+}\q_{[2211]}^{kjnlim} 
      \nonumber \\[6pt]
    &&\qquad\left.      {+}\q_{[2211]}^{jmknil}{-}\q_{[2211]}^{imknjl}
                         {-}\q_{[2211]}^{jmlnik}{+}\q_{[2211]}^{imlnjk}
                         {-}\q_{[2211]}^{jnkmil}{+}\q_{[2211]}^{inkmjl}
                         {+}\q_{[2211]}^{jnlmik}{-}\q_{[2211]}^{inlmjk}\right)\ .
\la{trfree}
\eea
These satisfy the useful identites
\bea
&& 
t_{[22]}^{jnim} +  t_{[22]}^{jmin} \equiv 0 
\\[8pt]
&& 
t_{[222]}^{jkmiln} +  t_{[222]}^{jmnilm} \equiv 0 
\\[8pt]
&& 
t_{[2]}^{ij} t_{[2]}^{km} - t_{[2]}^{kj} t_{[2]}^{im} 
                         - \fr12 t_{[22]}^{knip} t_{[22]}^{pjnm} \equiv 0
\\[8pt]
&& 
t_{[22]}^{minl} t_{[22]}^{lqjp} - t_{[22]}^{mjnl} t_{[22]}^{lqip} 
                + \fr12 t_{[22]}^{ikjl} t_{[222]}^{mkpnlq} \equiv 0
\\[8pt]
&& 
t_{[3]}^{ijk} t_{[3]}^{knm} - t_{[3]}^{njk} t_{[3]}^{kim} 
                + \fr{1}{2N} t_{[22]}^{npiq} t_{[22]}^{qjpm} \equiv 0
\\[8pt]
&& 
t_{[22]}^{kmin} t_{[22]}^{njmp} + t_{[22]}^{kmjn} t_{[22]}^{minp}
                                + t_{[22]}^{imjn} t_{[22]}^{nkmp} \equiv 0
\\[8pt]
&& 
t_{[222]}^{qimpjn} t_{[222]}^{nkrmls} + t_{[222]}^{qkmpln} t_{[222]}^{jmrins}
                               + t_{[222]}^{jkmiln} t_{[222]}^{qmspnr} \equiv 0
\ .\eea

\subsection{The supercommutators of the superderivations}

Using the invariant $\q$ and $t$ tensors, 
the most general permutation invariant and 
Lorentz covariant supercommutation relations of the
superderivations may be expressed, 
\bea
\{ \n^i_{\a}, \n^j_{\da} \}&=&
      \left(a_2t_{[2]}^{ij}+a_{11}\q_{[11]}^{ij}\right)\n_{\a\da}
              \la{a} \\[8pt]
\{ \n^i_{\a}, \n^j_{\b} \}&=&
       \left(b_2 t_{[2]}^{ij}+b_{11}\q_{[11]}^{ij}\right)M_{\a\b}
   +\e_{\a\b}b_{21} \left(t_{[21]}^{ikj}-t_{[21]}^{jki}\right)Y^{k}
         \nonumber\\
   &&+\e_{\a\b}b_{22} t_{[22]}^{ikjl} Z^{kl}
                  \la{b} \\[8pt]
\{ \n^i_{\da}, \n^j_{\db} \}&=&
       \left(\C{b}_2 t_{[2]}^{ij}+\C{b}_{11}\q_{[11]}^{ij}\right)M_{\da\db}
   +\e_{\da\db}\C{b}_{21} \left(t_{[21]}^{ikj}-t_{[21]}^{jki}\right)Y^{k}
         \nonumber\\
   &&+\e_{\da\db}\C{b}_{22} t_{[22]}^{ikjl} Z^{kl}
                \la{bb} \\[8pt]
 \left[ \n^i_{\a}, \n_{\b\db}  \right] &=&
     \e_{\a\b}\left(c_2 t_{[2]}^{ij}+c_{11}\q_{[11]}^{ij}\right)\n^j_{\db}
    \la{c} \\[8pt]
 \left[ \n^i_{\da}, \n_{\b\db} \right]&=&
    \e_{\da\db}\left(\C{c}_2 t_{[2]}^{ij}
    +\C{c}_{11}\q_{[11]}^{ij}\right)\n^j_{\b}
               \la{bc} \\[8pt]
 \left[ \n_{\a\da}, \n_{\b\db} \right]&=&
     r \left( \e_{\a\b}M_{\da\db}+ \e_{\da\db}M_{\a\b}\right) \ \ .
    \la{r} 
\eea
\goodbreak\noindent
Comments:

\noindent
a) The equations involving $Y^{i}$ and $Z^{ij}=-Z^{ji}$ 
on the right hand side have been written so as to exhibit 
manifestly the irreducibility of these operators. In particular, use of the 
partially trace-free 
invariant tensors as coefficients automatically yields  $Y^{i}$ satisfying \re{irYeq} 
and $Z^{ij}$ satisfying \re{irZeq}, since using these tensors
guarantees that the relevant term vanishes when one replaces 
$Z^{kl}$  by $\q^{k}_{[1]}V^l$ and independently $Y^{k}$ by $\q^{k}S$. 

\noindent
b) For the $\n_{\a}^i$ and the $\n_{\da}^i$, we have not separated 
the permutation-irreducible tensors explicitly. 
However, the tensors $t$ from \re{trfree} 
have been chosen to 
correspond to the decomposition into the irreducible pieces.

\noindent
c) That the two terms on the right-hand-side of \re{r} always have the same
coefficient, can be easily deduced from the 
Jacobi identity for three $\n_{\a\da}$'s.
The parameter $r$ distinguishes the two main classes of supersymmetry algebras
we shall consider: The contraction to the $r=0$ case corresponds to the 
{\it algebras of super-Poincar\'e type} and for $r\neq 0$ we obtain 
{\it algebras of super de Sitter type}.
We shall not consider algebras of superconformal type, which have a second
element transforming as a Lorentz vector, the generator of conformal
transformations.

\noindent
d) The right hand sides involve the most general Lorentz covariant
terms. This guarantees that Jacobi identities involving one $M$ are 
automatically satisfied.

\noindent
e) The fifteen complex parameters
$\{a_2,a_{11}\}$, $\{b_2,b_{11}, b_{21}, b_{22}\}$, 
$\{\C{b}_2,\C{b}_{11},\C{b}_{21},\C{b}_{22}\}$,
$\{c_2,c_{11}\}$, $\{\C{c}_2,\C{c}_{11}\}$, and $\{r\}$ are a priori independent.
They are to be chosen so as to satisfy the super Jacobi identities, which we shall
consider in the next section.

\subsection{The action of $\gh$ on the superderivations}

The most general commutation 
relations of the Lorentz scalar operators $Y$ and $Z$ with the 
superderivations compatible with Lorentz and permutation covariance, e.g.
\be
 \left[ Y^i,\n^j_{\a} \right] =
              \left(d_3 \q_{[3]}^{ijk}
                   +d_{21}^a \q_{[21]}^{ijk}
                   +d_{21}^b \q_{[21]}^{ikj}
                   +d_{21}^c\q_{[21]}^{kji}
                   +d_{111}\q_{[111]}^{ijk}\right)\n^k_{\a}\ ,
\label{YDelcom}
\ee
on imposition of the trace conditions, yield the following 
eight-parameter set of relations involving the 
partially trace-free tensors \re{trfree}: 
\bea
 \left[ Y^i,\n^j_{\a} \right]&=&
              \left(d_3 t_{[3]}^{ijk}
                   +d_{21}^a t_{[21]}^{ijk}
                   +d_{21}^b t_{[21]}^{ikj}\right)\n^k_{\a}
      \la{YS} \\[8pt]              
 \left[ Y^i,\n^j_{\da} \right]&=&
              \left(\C{d}_3 t_{[3]}^{ijk}
                   +\C{d}_{21}^a t_{[21]}^{ijk}
                   +\C{d}_{21}^b t_{[21]}^{ikj}\right)\n^k_{\da}
        \\[8pt]              
 \left[ Y^i,\n_{\a\da} \right]&=&0
        \\[8pt]             
 \left[ Z^{ij},\n^k_{\a} \right]&=&
       f_{22} t_{[22]}^{jkil} \n^l_{\a}
        \\[8pt]              
 \left[ Z^{ij},\n^k_{\da} \right] &=& \C{f}_{22} t_{[22]}^{jkil} \n^l_{\da}
        \\[8pt]              
 \left[ Z^{ij},\n_{\a\da} \right]&=&0 
 \la{ZD} \ \ .              
\eea
We note that the 
Coleman-Mandula type relation,  $[\gh , \cd_0 ] =0$,
is an immediate consequence of the trace conditions.
However, the internal symmetry can still act non-trivially
on the odd derivations.

\subsection{The commutators in $\gh$}

The subalgebra of the $Y$'s and $Z$'s has the Lorentz and 
permutation covariant form 
satisfying the trace-conditions:
\bea
 [ Y^i, Y^j ]&=&  g_{22} t_{[22]}^{imjn}Z^{mn}
 \la{yy}       \\[8pt]              
 [ Z^{ij},Y^k ]&=& h_{22} t_{[22]}^{jkil} Y^l 
                   + h_{32} t_{[32]}^{jklim}Z^{lm}
 \la{zy}      \\[8pt]                
 [ Z^{ij},Z^{kl} ]&=& k_{222} t_{[222]}^{jkmiln}Z^{mn}\ .
\la{zz}\eea
In fact the Jacobi identities always imply that $h_{32}=0$ (see below). 
This reduces the number of parameters to the three, which are 
constrained by the Jacobi identities.

\section{Democratic Lie algebras $\gg$}

The a priori Lorentz covariant commutators of our $N$-extended 
democratic algebras must satisfy super Jacobi identities which guarantee
that the products of the underlying operators are associative.
We shall now consider the constraints imposed on the parameters in 
\re{a}-\re{r}, \re{YS}-\re{ZD},\re{yy}-\re{zz} by the super Jacobi identities. 
Let us first recall that, by construction,
all the Jacobi identities involving at least one $M$ are
automatically satisfied.
We begin with the subalgebra $\gh$.

\subsection{Democratic Lie algebras $\gh$}\la{ghlist}

To find all possible $S_N$ democratic algebras containing the $N(N{-}1)/2$ 
generators $Y$ and $Z$, the Jacobi identities for \re{yy}-\re{zz} 
need to be satisfied. 
These yield the following four conditions on the 
four parameters $g_{22},h_{22},h_{32},k_{222}$:
\bea
h_{32}h_{22}\ =\
h_{32}k_{222}&=&0
       \nonumber\\
h_{22}(h_{22}-2 k_{222})&=&0
       \nonumber\\
N g_{22}(h_{22}-2 k_{222})-2 h_{32}^2&=&0 \ \ .
\eea
They generally imply that $\, h_{32}=0\, $, leaving the conditions
\bea
h_{22}(h_{22}-2 k_{222})&=&0
       \nonumber\\
g_{22}(h_{22}-2 k_{222})&=&0\ .
\eea
These equations lead to a classification in five distinct categories:

\noindent
{\bf 1a.\ } Abelian $\gh$: all the scalar operators commute
\be
g_{22}=h_{22}=k_{222}=0 
\la{comscalars}
\ee
and the $Y$ and $Z$ can still be renormalised freely.

\noindent
{\bf 1b.\ } The $Z$'s commute, they commute with the $Y$'s 
but the commutators of the $Y$'s generate the $Z$'s. 
By renormalisation of the $Z$'s or the $Y$'s, we find
\be
g_{22}= 1\quad,\quad h_{22}=k_{222}=0 \ \ .
\la{ygenerated}
\ee

\noindent
{\bf 2.\ }  The $Z$'s form an $so(N{-}1)$ algebra, 
with $N{-}1$ commuting $Y$'s which moreover are $so(N{-}1)$ scalars, 
i.e. do not transform under the $Z$. 
Using the normalisation freedom, we may write
\be
g_{22}=h_{22}=0 \quad,\quad k_{222}= 1\ \ .
\la{sonminus}
\ee

\noindent
{\bf 3a.\ }  The inhomogeneous $so(N{-}1)$ case. 
By normalisation of the $Z$'s, the parameters can be brought to
\be
g_{22}=0\quad,\quad h_{22}=2\quad,\quad k_{222}=1\ \ .
\la{inhson}
\ee
The $Y$'s behave as a vector under $so(N{-}1)$ and commute. 
They behave as momenta with respect to $so(N{-}1)$ and hence this 
corresponds to an inhomogeneous $so(N{-}1)$ algebra. 
The normalisations of the $Y$'s can still be adjusted freely. 

\noindent
{\bf 3b.\ } The $so(N)$ case.
We clearly have as many $Y$ and $Z$ operators as there are generators of $so(N)$,
which is indeed a particular democratic Lie algebra $\gh$.
In this case, by suitable renormalisations of the $Y$'s and the $Z$'s, 
the parameters can be brought to their $so(N)$ values, which we normalise as:
\be
g_{22}=1\quad,\quad h_{22} = 2 \quad,\quad k_{222} = 1\ .  
   \la{sonpargen}
\ee
That these values correspond to $so(N)$ can be seen as follows.
The commutation relations of the $N(N{-}1)/2$ generators 
$M^{ij}=M^{ji}$ of $so(N)$ are usually written as,
$$
[ M^{ij}, M^{kl} ] = \q^{jk}_{[2]} M^{il} - \q^{ik}_{[2]} M^{jl} 
                     - \q^{jl}_{[2]} M^{ik} + \q^{il}_{[2]} M^{jk}\ .
$$
Defining projections
\bea
V^{j}&=&\q^{k}_{[1]} M^{kj}
   \nonumber\\
T^{jk}&=&M^{jk}-\fr{1}{N}\left(\q^{j}_{[1]}V^{k} \q^{k}_{[1]}V^{j}\right)
\ \  ,
\la{democson}
\eea
we obtain that the subset of the $T$ operators alone form a democratic 
$so(N{-}1)$ subalgebra  (with $(N{-}1)(N{-}2)/2$ 
independent operators) of the $so(N)$ algebra.  
The $N{-}1$ independent $V$ operators transform as a vector 
under the $so(N{-}1)$ subalgebra.
The $V$ and the $T$ satisfy precisely the commutation 
relations \re{yy}-\re{zz} satisfied by
$Y$ and $Z$ respectively with
\be
g_{22}\ =\ -\fr{N}{2} \quad,\quad
h_{22}\ =\ 1          \quad,\quad
k_{222}\ =\  \fr{1}{2} \ \ .
\la{sonpar}
\ee
Since there are possible arbitrary democratic rescalings of $V$ with respect to $Y$ and of $T$ with respect to $Z$, the algebra of the $Y$'s 
and the $Z$'s corresponds to an
$so(N)$ algebra provided \re{sonpargen} holds.

\subsection{Supersymmetry algebras $\gg$}

The full discussion for the rest of the super Jacobi identities 
is rather intricate. 
We discuss the full set of solutions in the appendix, 
discussing the main features here. 

We have chosen to discuss the general solution of the Jacobi identities 
in terms of two criteria:
\begin{enumerate}
\item the first criterion is related to the appearance
of the term $\q_{[2]}^{ij}\n_{\a\da}$ 
in the anticommutators of $\n_{\a}^i$ with $\n_{\da}^j$ 
(parameter $a_{2}$) and of the $Y$'s in the 
anticommutator of two $\n_{a}$'s 
(parameter $b_{21}$) or of two $\n_{\da}$'s (parameter $\C{b}_{21}$).
\item the second criterion reveals the structure of the algebra $\gh$ of the 
Lorentz scalar elements as discussed in the preceding section.
\end{enumerate}
We use the values of the parameters $a_2$, $b_{21}$ and $\C{b}_{21}$ 
as the basis of our classification. It follows from from 
\re{a}, \re{b} and \re{bb} that, if any of these three parameters 
is non zero, it may be renormalised to one by rescaling  the three superderivations democratically. 
Hence, using also the fact that we have a natural symmetry under 
the interchange of the dotted and undotted operators, we are led 
to six independent classes of superalgebras:
\bea
{\rm{Class\ A\ :}}&& a_2=1,\  b_{21}=1,\ \C{b}_{21}=1
        \nonumber\\
{\rm{Class\ B\ :}}&& a_2=1,\  b_{21}=1,\ \C{b}_{21} =0
        \nonumber\\
{\rm{Class\ C\ :}}&& a_2=0 ,\  b_{21}=1,\ \C{b}_{21}=1
        \nonumber\\
{\rm{Class\ D\ :}}&& a_2=1,\  b_{21}=0 ,\ \C{b}_{21}=0 
        \nonumber\\
{\rm{Class\ E\ :}}&& a_2=0 ,\  b_{21}=1,\ \C{b}_{21}=0 
        \nonumber\\
{\rm{Class\ F\ :}}&& a_2=0 ,\  b_{21}=0 ,\ \C{b}_{21}=0 
        \la{cases}
\eea
which we discuss in detail in the Appendix. 
Classes $B$ and $E$ are {\it chiral}, not having the mirror symmetry under
the chirality interchanges between dotted and undotted indices 
($\a \leftrightarrow \da,\ldots$) and between the parameters 
$c \leftrightarrow \C{c},\ldots$ (for existing unbarred-barred pairs). 
The two further classes
\bea
{\rm{Class\ B'\ :}}&& a_2=1,\  b_{21}=0 ,\ \C{b}_{21}=1
        \nonumber\\
{\rm{Class\ E'\ :}}&& a_2=0 ,\  b_{21}=0 ,\ \C{b}_{21}=1
        \la{primecases}
\eea
can clearly be obtained trivially from the $B$ and $E$ classes by performing 
the above chirality exchanges; and we do not explicitly discuss these.

Within the above classes, the discussion is subdivided according to the 
values of $k_{222}$ and $h_{22}$, corresponding to
the division in section \ref{ghlist},
\bea
{\rm{Case\ 1\ :}}&& k_{222}=0,\  h_{22}=0
        \nonumber\\
{\rm{Case\ 2\ :}}&& k_{222}=1,\  h_{22} =0
        \nonumber\\
{\rm{Case\ 3\ :}}&& k_{222}=1,\   h_{22}=2 \ \ .
        \la{subcases}
\eea

\subsection{Some solutions of the super Jacobi Identities}\la{examples}
In this section, we discuss the main noteworthy features revealed by our approach.
Let us consider
Case A3 from the appendix:
\bea
\{ \n^i_{\a}, \n^j_{\da} \}&=&
      \left(
        t_{[2]}^{ij}
        +a_{11}\q_{[11]}^{ij}\right)\n_{\a\da}
              \la{A3a} \\[5pt]
\{ \n^i_{\a}, \n^j_{\b} \}&=&
      4  
      \left(b_{22} t_{[2]}^{ij}
      + a_{11} \C{b}_{22}\q_{[11]}^{ij}\right)M_{\a\b} 
              \nonumber\\
   && +\e_{\a\b} \left(\left(t_{[21]}^{ikj}-t_{[21]}^{jki}\right)Y^{k}
        + b_{22} t_{[22]}^{ikjl} Z^{kl} \right)
              \la{A3b} \\[5pt]
\{ \n^i_{\da}, \n^j_{\db} \}&=&
      4  
       \left(\C{b}_{22} t_{[2]}^{ij}
     + a_{11}b_{22}\q_{[11]}^{ij}\right)M_{\da\db}
             \nonumber\\
   &&  +\e_{\da\db}\left( \left(t_{[21]}^{ikj}-t_{[21]}^{jki}\right)Y^{k}
         +\C{b}_{22} t_{[22]}^{ikjl} Z^{kl} \right)
              \la{A3bb} \\[5pt]
\left[ \n^i_{\a}, \n_{\b\db}  \right] &=&
    4 \e_{\a\b} \left(b_{22} t_{[2]}^{ij}
         +\fr{\C{b}_{22}}{N}\q_{[11]}^{ij}\right)\n^j_{\db}
             \la{A3c} \\[5pt]
\left[ \n^i_{\da}, \n_{\b\db} \right]&=&
    4 \e_{\da\db}\left(\C{b}_{22} t_{[2]}^{ij}
     +\fr{b_{22}}{N} \q_{[11]}^{ij}\right)\n^j_{\b}
            \la{A3bc} \\[5pt]
\left[ \n_{\a\da}, \n_{\b\db} \right]&=&
     16b_{22}\C{b}_{22} \left( \e_{\a\b}M_{\da\db}
     + \e_{\da\db}M_{\a\b}\right) 
             \la{A3r}  \\[10pt]
\left[ Y^i,\n_{\a\da} \right]&=&0
                 \\[5pt]             
\left[ Z^{ij},\n_{\a\da} \right]&=&0 
                 \la{A3ZD}\\[5pt]   
\left[ Y^i,\n^j_{\a} \right]&=&  
             4  \left(
              \C{b}_{22} a_{11} t_{[21]}^{ikj}
                -\fr{b_{22}}{N} t_{[21]}^{ijk}
                 \right)\n^k_{\a} 
               \la{A3YS} \\[5pt]              
\left[ Y^i,\n^j_{\da} \right]&=&
              4  \left(
               b_{22}  a_{11} t_{[21]}^{ikj}
              -\fr{\C{b}_{22}}{N} t_{[21]}^{ijk}
              \right)\n^k_{\da}
                \\[5pt]              
\left[ Z^{ij},\n^k_{\a} \right]&=& 2 t_{[22]}^{jkil} \n^l_{\a}
                 \\[5pt]              
\left[ Z^{ij},\n^k_{\da} \right] &=& 2 t_{[22]}^{jkil} \n^l_{\da}
                \\[5pt]
[ Y^i, Y^j ]&=&  -4 a_{11}b_{22}\C{b}_{22} t_{[22]}^{imjn}Z^{mn}
                \\[5pt]              
[ Z^{ij},Y^k ]&=& 2 t_{[22]}^{jkil} Y^l
               \\[5pt]                
[ Z^{ij},Z^{kl} ]&=&  t_{[222]}^{jkmiln}Z^{mn}\ \ .
\eea
The main unusual features displayed by this algebra are:
\begin{description}
\item{1)} non-trivial action of the subalgebra $\gh$ on the vector space
of superderivations $\cd$,
\item{2)} nonabelian subalgebra of the Lorentz scalar generators,
\item{3)} occurrence of the $a_{11}$ term in \re{A3a}.
\end{description}

The above example is of super de Sitter type.
A chiral super Poincar\'e type example, also displaying these interesting features, is given by Case B3:
\bea
\{ \n^i_{\a}, \n^j_{\da} \}&=&
      \left
      (t_{[2]}^{ij}
       +a_{11}\q_{[11]}^{ij}\right)\n_{\a\da}
                \la{B3a} \\[5pt]
\{ \n^i_{\a}, \n^j_{\b} \}&=&
       \left( 4 b_{22}  t_{[2]}^{ij}
             +N c_{11} a_{11}\q_{[11]}^{ij}    
      \right)M_{\a\b} \nonumber\\[6pt]
     && +\e_{\a\b} \left(\left(t_{[21]}^{ikj}-t_{[21]}^{jki}\right)Y^{k}
        + b_{22} t_{[22]}^{ikjl} Z^{kl} \right)
                \la{B3b} \\[5pt]
\{ \n^i_{\da}, \n^j_{\db} \}&=& 0
                  \la{B3bb} \\[5pt]
\left[ \n^i_{\a}, \n_{\b\db}  \right] &=&
     \e_{\a\b}\left( 4 b_{22}  t_{[2]}^{ij}
      +c_{11}\q_{[11]}^{ij}\right)\n^j_{\db}
                   \la{B3c}  \\[5pt]
\left[ \n^i_{\da}, \n_{\b\db} \right]&=&0    
                   \la{B3bc} \\[5pt]
\left[ \n_{\a\da}, \n_{\b\db} \right]&=&0    
                   \la{B3r}  \\[10pt]
\left[ Y^i,\n_{\a\da} \right]&=&0
                            \\[5pt]             
\left[ Z^{ij},\n_{\a\da} \right]&=&0 
                     \la{B3ZD}\\[5pt]   
\left[ Y^i,\n^j_{\a} \right]&=&
              \left( -\fr{4}{N} b_{22}  t_{[21]}^{ijk}
                        +N a_{11} c_{11} t_{[21]}^{ikj} \right)\n^k_{\a}
                     \la{B3YS} \\[5pt]              
\left[ Y^i,\n^j_{\da} \right]&=&
              \left( - c_{11} t_{[21]}^{ijk} 
           +4 a_{11} b_{22} t_{[21]}^{ikj}\right)
              \n^k_{\da}
                        \\[5pt]              
\left[ Z^{ij},\n^k_{\a} \right]&=& 2 t_{[22]}^{jkil} \n^l_{\a}
                        \\[5pt]              
\left[ Z^{ij},\n^k_{\da} \right] &=& 2 t_{[22]}^{jkil} \n^l_{\da}
                         \\[5pt]              
[ Y^i, Y^j ]&=& 
            -N a_{11} c_{11} b_{22} t_{[22]}^{imjn}Z^{mn}
                              \\[5pt]              
[ Z^{ij},Y^k ]&=& 2 t_{[22]}^{jkil} Y^l
                            \\[5pt]                
[ Z^{ij},Z^{kl} ]&=&  t_{[222]}^{jkmiln}Z^{mn}\ .
\eea

\section{Conclusion}

The inclusion of multiplicities in our programme \cite{DN}, extending in a Lorentz covariant way the algebra of coordinates and derivatives, has been shown to exhibit interesting new features and a rather rich structure of solutions for the super-Jacobi identities.
In order to obtain explicit solutions, we have chosen 
to restrict ourselves in this article to a set of 
operators of spin less than or equal to one and 
to impose {\it{democracy}}.
Within these restricted hypotheses, 
we have been able to classify fully the allowed 
superalgebras of derivations and superderivations. 
Apart from the well-known examples \cite{HLS,N}, new and 
potentially interesting cases have been uncovered. 

\vskip 0.5  true cm
\noindent
{\large\bf{Acknowledgements}}

\noindent
This work has been supported in part by the TMR European Network
``Integrability, non-pertur\-bative effects and symmetry in
quantum field theory'' (contract number FMRX-CT96-0012). One of the authors 
(J.N.) wishes to thank the Theory Division at CERN where part of this work was
carried through in April 2000 and the other (C.D.) thanks the MPI f\"ur 
Mathematik, Bonn and the Sonderforschungsbereich SFB 256, Mathematisches 
Institut der Universit\"at Bonn for hospitality during the performance
of this work. 

\vfil \newpage

\appendix
\noindent
{\Large {\bf {Appendix}}}      

\vskip 5pt
\noindent
With classes (A-F) defined in \re{cases} 
and subcases (1-3) defined by \re{subcases}
the full classification of the democratic supersymmetry algebras 
is given below. 
\renewcommand\thesection{Class \Alph{section}}
\section{}    
\setcounter{equation}{0}
\renewcommand\theequation{\Alph{section}\arabic{equation}}

Imposing the super Jacobi identities together with the class A constraints,
$a_2=1,\, b_{21}=1,\,\C{b}_{21}=1$,
yields the relations
\bea
\{ \n^i_{\a}, \n^j_{\da} \}&=&
      \left(
        t_{[2]}^{ij}
        +a_{11}\q_{[11]}^{ij}\right)\n_{\a\da}
              \la{Aa} \\[5pt]
\{ \n^i_{\a}, \n^j_{\b} \}&=&
      4 k_{222} 
      \left(b_{22} t_{[2]}^{ij}
      + a_{11} \C{b}_{22}\q_{[11]}^{ij}\right)M_{\a\b} 
              \nonumber\\
   && +\e_{\a\b} \left(\left(t_{[21]}^{ikj}-t_{[21]}^{jki}\right)Y^{k}
        + b_{22} t_{[22]}^{ikjl} Z^{kl} \right)
              \la{Ab} \\[5pt]
\{ \n^i_{\da}, \n^j_{\db} \}&=&
      4 k_{222} 
       \left(\C{b}_{22} t_{[2]}^{ij}
     + a_{11}b_{22}\q_{[11]}^{ij}\right)M_{\da\db}
             \nonumber\\
   &&  +\e_{\da\db}\left( \left(t_{[21]}^{ikj}-t_{[21]}^{jki}\right)Y^{k}
         +\C{b}_{22} t_{[22]}^{ikjl} Z^{kl} \right)
              \la{Abb} \\[5pt]
\left[ \n^i_{\a}, \n_{\b\db}  \right] &=&
    4 \e_{\a\b} k_{222}\left(b_{22} t_{[2]}^{ij}
         +\fr{\C{b}_{22}}{N}\q_{[11]}^{ij}\right)\n^j_{\db}
             \la{Ac} \\[5pt]
\left[ \n^i_{\da}, \n_{\b\db} \right]&=&
    4 \e_{\da\db}k_{222}\left(\C{b}_{22} t_{[2]}^{ij}
     +\fr{b_{22}}{N} \q_{[11]}^{ij}\right)\n^j_{\b}
            \la{Abc} \\[5pt]
\left[ \n_{\a\da}, \n_{\b\db} \right]&=&
     16b_{22}\C{b}_{22} k_{222}^2 \left( \e_{\a\b}M_{\da\db}
     + \e_{\da\db}M_{\a\b}\right) 
             \la{Ar}  \\[10pt]
\left[ Y^i,\n_{\a\da} \right]&=&0
                 \\[5pt]             
\left[ Z^{ij},\n_{\a\da} \right]&=&0 
                 \la{AZD}\\[5pt]   
\left[ Y^i,\n^j_{\a} \right]&=&  
             4 k_{222} \left(
              \C{b}_{22} a_{11} t_{[21]}^{ikj}
                -\fr{b_{22}}{N} t_{[21]}^{ijk}
                 \right)\n^k_{\a} 
               \la{AYS} \\[5pt]              
\left[ Y^i,\n^j_{\da} \right]&=&
              4 k_{222} \left(
               b_{22}  a_{11} t_{[21]}^{ikj}
              -\fr{\C{b}_{22}}{N} t_{[21]}^{ijk}
              \right)\n^k_{\da}
                \\[5pt]              
\left[ Z^{ij},\n^k_{\a} \right]&=& h_{22} t_{[22]}^{jkil} \n^l_{\a}
                 \\[5pt]              
\left[ Z^{ij},\n^k_{\da} \right] &=& h_{22} t_{[22]}^{jkil} \n^l_{\da}
                \\[5pt]
[ Y^i, Y^j ]&=&  -4 a_{11}b_{22}\C{b}_{22}k_{222} t_{[22]}^{imjn}Z^{mn}
                \\[5pt]              
[ Z^{ij},Y^k ]&=& h_{22} t_{[22]}^{jkil} Y^l
               \\[5pt]                
[ Z^{ij},Z^{kl} ]&=& k_{222} t_{[222]}^{jkmiln}Z^{mn}\ \ ,
\eea
with the space of class A superalgebras defined 
by solutions of the system of quadratic equations
\be\arr
b_{22}(2k_{222}-h_{22})&=&0 \\
\C{b}_{22}(2k_{222}-h_{22})&=&0 \\
h_{22}(2k_{222}-h_{22})&=&0\ . 
\ea\ee
We find three subcases (see \re{subcases})

\begin{description}     
\item{\underline{Case A1}}
Since $h_{22}=k_{222}=0$,
the parameters $a_{11}$, $b_{22}$ and $\C{b}_{22}$  are free.
This includes the standard \sp\ with abelian algebra $\gh$
of central charges. 

\item{\underline{Case A2}}
Here $a_{11}$ is free, $k_{222}=1$ and all other parameters are zero.
There is an $so(N{-}1)$ subalgebra (see \re{sonminus}) of the $Z$'s which 
decouples. 

\item{\underline{Case A3}}
This is a much less trivial case (see section \ref{examples}) and 
the full $so(N)$ algebra \re{sonpar} is included in the algebra. 
The independent parameters are 
$a_{11}$, $b_{22}$, $\C{b}_{22}$ while $h_{22}=2, k_{222}=1$. 
\end{description}                                   

\section{}\setcounter{equation}{0}    

This class is {\it chiral} of super Poincar\'e type: $a_2=1,\  b_{21}=1,\ \C{b}_{21}=0$.
It has relations
\bea
\{ \n^i_{\a}, \n^j_{\da} \}&=&
      \left
      (t_{[2]}^{ij}
       +a_{11}\q_{[11]}^{ij}\right)\n_{\a\da}
                \la{Ba} \\[5pt]
\{ \n^i_{\a}, \n^j_{\b} \}&=&
       \left( 4 b_{22} k_{222}  t_{[2]}^{ij}
             +N c_{11} a_{11}\q_{[11]}^{ij}    
      \right)M_{\a\b} \nonumber\\[6pt]
     && +\e_{\a\b} \left(\left(t_{[21]}^{ikj}-t_{[21]}^{jki}\right)Y^{k}
        + b_{22} t_{[22]}^{ikjl} Z^{kl} \right)
                \la{Bb} \\[5pt]
\{ \n^i_{\da}, \n^j_{\db} \}&=& 
       \e_{\da\db}\C{b}_{22} t_{[22]}^{ikjl} Z^{kl}
                  \la{Bbb} \\[5pt]
\left[ \n^i_{\a}, \n_{\b\db}  \right] &=&
     \e_{\a\b}\left( 4 b_{22} k_{222} t_{[2]}^{ij}
      +c_{11}\q_{[11]}^{ij}\right)\n^j_{\db}
                   \la{Bc}  \\[5pt]
\left[ \n^i_{\da}, \n_{\b\db} \right]&=&0    
                   \la{Bbc} \\[5pt]
\left[ \n_{\a\da}, \n_{\b\db} \right]&=&0    
                   \la{Br}  \\[10pt]
\left[ Y^i,\n_{\a\da} \right]&=&0
                            \\[5pt]             
\left[ Z^{ij},\n_{\a\da} \right]&=&0 
                     \la{BZD}\\[5pt]   
\left[ Y^i,\n^j_{\a} \right]&=&
              \left( -\fr{4}{N} b_{22} k_{222} t_{[21]}^{ijk}
                        +N a_{11} c_{11} t_{[21]}^{ikj} \right)\n^k_{\a}
                     \la{BYS} \\[5pt]              
\left[ Y^i,\n^j_{\da} \right]&=&
              \left( - c_{11} t_{[21]}^{ijk} 
           +4 a_{11} b_{22}k_{222} t_{[21]}^{ikj}\right)
              \n^k_{\da}
                        \\[5pt]              
\left[ Z^{ij},\n^k_{\a} \right]&=& h_{22} t_{[22]}^{jkil} \n^l_{\a}
                        \\[5pt]              
\left[ Z^{ij},\n^k_{\da} \right] &=& h_{22} t_{[22]}^{jkil} \n^l_{\da}
                         \\[5pt]              
[ Y^i, Y^j ]&=& 
            -N a_{11} c_{11} b_{22} t_{[22]}^{imjn}Z^{mn}
                              \\[5pt]              
[ Z^{ij},Y^k ]&=& h_{22} t_{[22]}^{jkil} Y^l
                            \\[5pt]                
[ Z^{ij},Z^{kl} ]&=& k_{222} t_{[222]}^{jkmiln}Z^{mn}\ \ .
\eea
Here the parameters are constrained by the system of equations
\be\arr
\C{b}_{22}h_{22}\ =\
\C{b}_{22} k_{222}&=&0                      \\
b_{22}(2k_{222}-h_{22})&=&0                 \\
h_{22}(2k_{222}-h_{22})&=&0 
\ea\ee
defining the space of class B superalgebras. They are all of
chiral super-Poincar\'e type.
There are three subcases of solutions (see \re{subcases})

\begin{description}                                
\item{\underline{Case B1}}
The parameters
$a_{11}$, $b_{22}$, $\C{b}_{22}$ and $c_{11}$ are free, 
$h_{22}=k_{222}=0$. 
The $Z$'s are central, not the $Y$'s.

\item{\underline{Case B2}}
The parameters  $a_{11}$, and $c_{11}$  
are free, $k_{222} =1$ and the remaining are zero.
The subalgebra $\gh$ contains the $so(N{-}1)$ of the $Z$'s which decouples. The subalgebra of the $Y$'s is abelian.  

\item{\underline{Case B3}}
The parameters  $a_{11}$,  $b_{22}$, and $c_{11}$  
are free,  $h_{22}=2, k_{222}=1, \C{b}_{22}=0$ (see \ref{examples}).
\end{description}                            

\section{}\setcounter{equation}{0}   

This class contains super algebras  of de Sitter type. They allow contractions to super Poincar\'e type algebras by setting $c_2$ and/or $\C{c}_2$ to zero.
The relations $a_2=0 ,\  b_{21}=1,\ \C{b}_{21}=1$ yield the
superbrackets
\bea
\{ \n^i_{\a}, \n^j_{\da} \}&=& a_{11}\q_{[11]}^{ij}\n_{\a\da}
                   \la{Ca} \\[5pt]
\{ \n^i_{\a}, \n^j_{\b} \}&=&    a_{11} \C{c}_2 \q_{[11]}^{ij} M_{\a\b}
   +\e_{\a\b}\left( \left(t_{[21]}^{ikj}-t_{[21]}^{jki}\right)Y^{k}
          + b_{22} t_{[22]}^{ikjl} Z^{kl} \right)
                       \la{Cb} \\[5pt]
\{ \n^i_{\da}, \n^j_{\db} \}&=&  a_{11}  c_2 \q_{[11]}^{ij} M_{\da\db}
   +\e_{\da\db}\left( \left(t_{[21]}^{ikj}-t_{[21]}^{jki}\right)Y^{k}
          + \C{b}_{22} t_{[22]}^{ikjl} Z^{kl} \right)
                     \la{Cbb} \\[5pt]
\left[ \n^i_{\a}, \n_{\b\db}  \right] &=&
     \e_{\a\b}\left(c_2 t_{[2]}^{ij}
                    +\fr{\C{c}_2}{N} \q_{[11]}^{ij}\right)\n^j_{\db}
                 \la{Cc} \\[5pt]
\left[ \n^i_{\da}, \n_{\b\db} \right]&=&
    \e_{\da\db}\left(\C{c}_2 t_{[2]}^{ij}
                     + \fr{c_2}{N}\q_{[11]}^{ij}\right)\n^j_{\b}
                 \la{Cbc} \\[5pt]
\left[ \n_{\a\da}, \n_{\b\db} \right]&=&
      c_2 \C{c}_2 \left( \e_{\a\b}M_{\da\db}+ \e_{\da\db}M_{\a\b}\right) 
                     \la{Cr}  \\[10pt]
\left[ Y^i,\n_{\a\da} \right]&=&0
                               \\[5pt]             
\left[ Z^{ij},\n_{\a\da} \right]&=&0 \la{CZD}\\[5pt]   
\left[ Y^i,\n^j_\a \right]&=& a_{11} \C{c}_2  t_{[21]}^{ikj}\n^k_\a
                        \la{CYS} \\[5pt]              
\left[ Y^i,\n^j_{\da} \right]&=& a_{11} c_2  t_{[21]}^{ikj}\n^k_\da
                      \\[5pt]              
\left[ Z^{ij},\n^k_{\a} \right]&=& h_{22} t_{[22]}^{jkil} \n^l_{\a}
                         \\[5pt]              
\left[ Z^{ij},\n^k_{\da} \right] &=& h_{22} t_{[22]}^{jkil} \n^l_{\da}
                              \\[5pt]              
[ Y^i, Y^j ]&=&  -a_{11} b_{22} \C{c}_2 
         t_{[22]}^{imjn}Z^{mn}
                         \\[5pt]              
[ Z^{ij},Y^k ]&=& h_{22} t_{[22]}^{jkil} Y^l
       \\[5pt]                
 [ Z^{ij},Z^{kl} ]&=& k_{222} t_{[222]}^{jkmiln}Z^{mn}\ \ .
\eea
Here the parameters are constrained by the system of equations
\be\arr
\C{b}_{22} c_2 - b_{22} \C{c}_2 &=& 0                   \\ 
b_{22}h_{22}\ =\  b_{22} k_{222}&=&0                    \\
\C{b}_{22}h_{22}\ =\  \C{b}_{22} k_{222}&=&0            \\
a_{11}\C{c}_{2} b_{22}(2k_{222}-h_{22})&=&0             \\
h_{22}(2k_{222}-h_{22})&=&0 \ \ .
\ea\ee
We find three subcases (see \re{subcases})
\begin{description}                       
\item{\underline{Case C1}}
We have $h_{22}=k_{222}=0$ while
$a_{11}$ is free and 
$b_{22}$, $\C{b}_{22}$, $c_{2}$, $\C{c}_{2}$,
are constrained by the condition
\be
\C{b}_{22} c_{2}=b_{22}\C{c}_{2}\ \ .
\la{condC1}
\ee
The $Z$'s are central charges.

\item{\underline{Case C2}}
 The parameters $a_{11}$,  
$c_{2}$, $\C{c}_{2}$ are free, $k_{222}=1$ and the remaining are
zero. The subalgebra $so(N{-}1)\subset \gh$ of the $Z$'s decouples. 

\item{\underline{Case C3}}
The parameters 
 $a_{11}$,  $c_{2}$, $\C{c}_{2}$  are free, $h_{22}=2$, $k_{222}=1$ and 
$ b_{22}=\C{b}_{22}=0$. 
 
\end{description}   

\section{}\setcounter{equation}{0}   

This has $a_2=1,\  b_{21}=0 ,\ \C{b}_{21}=0 $, yielding 
\bea
\{ \n^i_{\a}, \n^j_{\da} \}&=&
      \left( t_{[2]}^{ij}+a_{11}\q_{[11]}^{ij}\right)\n_{\a\da}
                      \la{Da} \\[5pt]
\{ \n^i_{\a}, \n^j_{\b} \}&=& \e_{\a\b}b_{22} t_{[22]}^{ikjl} Z^{kl}
                      \la{Db} \\[5pt]
\{ \n^i_{\da}, \n^j_{\db} \}&=& \e_{\da\db}\C{b}_{22} t_{[22]}^{ikjl} Z^{kl}
                        \la{Dbb} \\[5pt]
\left[ \n^i_{\a}, \n_{\b\db}  \right] &=&0   \la{Dc} \\[5pt]
\left[ \n^i_{\da}, \n_{\b\db} \right]&=&0    \la{Dbc} \\[5pt]
\left[ \n_{\a\da}, \n_{\b\db} \right]&=&0   \la{Dr}  \\[10pt]
\left[ Y^i,\n_{\a\da} \right]&=&0
                                                  \\[5pt]             
\left[ Z^{ij},\n_{\a\da} \right]&=&0        \la{DZD}\\[5pt]   
 \left[ Y^i,\n^j_{\a} \right]&=&
              \left( d_3 t_{[3]}^{ijk} +d_{21}^a t_{[21]}^{ijk}
            - N \C{d}_{21}^a a_{11} t_{[21]}^{ikj}
                                            \right)\n^k_{\a}
                          \la{DYS} \\[5pt]              
\left[ Y^i,\n^j_{\da} \right]&=&
              \left( -d_3 t_{[3]}^{ijk} +\C{d}_{21}^a t_{[21]}^{ijk}
          - N a_{11} d_{21}^a t_{[21]}^{ikj}  
                        \right)\n^k_{\da}
                                \\[5pt]              
\left[ Z^{ij},\n^k_{\a} \right]&=& f_{22} t_{[22]}^{jkil} \n^l_{\a}
                              \\[5pt]              
\left[ Z^{ij},\n^k_{\da} \right] &=&  f_{22} t_{[22]}^{jkil} \n^l_{\da}
                               \\[5pt]              
[ Y^i, Y^j ]&=&  g_{22} t_{[22]}^{imjn}Z^{mn}
                               \\[5pt]              
[ Z^{ij},Y^k ]&=& h_{22} t_{[22]}^{jkil} Y^l
                           \\[5pt]                
[ Z^{ij},Z^{kl} ]&=& k_{222} t_{[222]}^{jkmiln}Z^{mn}\ \ .
\eea
The remaining parameters must satisfy the eighteen equations
\bea
&&d_3 b_{22}=0\ ,\ d_3 \C{b}_{22}=0
             \ ,\ d_3 f_{22}=0\ ,\ d_3 h_{22}=0
                             \\[5pt]
&&b_{22} f_{22}=0\ ,\ b_{22} h_{22}=0
           \ ,\ b_{22} k_{222}=0\ ,\ b_{22} a_{11} \C{d}_{21}^a=0 
                             \\[5pt]
&&\C{b}_{22} f_{22}=0\ ,\ \C{b}_{22} h_{22}=0
           \ ,\ \C{b}_{22} k_{222}=0\ ,\ \C{b}_{22} a_{11} d_{21}^a=0 
                             \\[5pt]
&&h_{22}(h_{22}-2k_{222})=0
              \ ,\ g_{22}(h_{22}-2k_{222})=0
           \ ,\ f_{22}(f_{22}-2k_{222})=0
                             \\[5pt]
&&d_{21}^a(f_{22}-h_{22})=0
              \ ,\ \C{d}_{21}^a(f_{22}-h_{22})=0
           \ ,\ d_3^2+2Nf_{22}g_{22}+N^3 a_{11}d_{21}^a\C{d}_{21}^a=0\ .
\eea
We find seven essentially different subcases (see \re{subcases})

\begin{description}   
\item{\underline{Case D1a}}
The parameter
$g_{22}$ is free while $h_{22}=k_{222}=b_{22}=\C{b}_{22}=f_{22}=0$ and 
$a_{11}$,  $d_{3}$, 
$d_{21}^a$, $\C{d}_{21}^a$, satisfy the condition
\be
     a_{11}d^a_{21}\C{d}^a_{21}+\fr{d_3^2}{N^3}=0\ .
\la{condD1a}
\ee

\item{\underline{Case D1b}}
The parameters
$b_{22}\neq 0$, $d_{21}^a$ and $g_{22}$  are free,
$a_{11}$ and $\C{d}_{21}^a$ are constrained by
\be
\C{d}_{21}^a a_{11} = 0
\la{condD1b}
\ee
and the remaining parameters are zero.

\item{\underline{Case D1c}}
The parameters 
$b_{22}\neq 0$, 
$\C{b}_{22}\neq 0$,  $g_{22}$ are free, 
$a_{11}$, $d_{21}^a$ , $\C{d}_{21}^a$
satisfy the conditions
\bea
{d}_{21}^a a_{11}& =& 0
       \nonumber\\
\C{d}_{21}^a  a_{11}& =& 0 
\la{condD1c}
\eea
and the remaining parameters are zero.

\item{\underline{Case D2a}}
Here $k_{222}=1$, the parameters $a_{11}$,  
$d_{3}$, $d_{21}^a$, $\C{d}_{21}^a$ satisfy the condition
\re{condD1a} and the remaining parameters are zero.

\item{\underline{Case D2b}}
All the parameters are zero except $k_{222}=1,\ f_{22}=2$ 
and $a_{11}$  
which is free.

\item{\underline{Case D3a}}
All the other parameters are zero except  $k_{222}=1,\ h_{22}=2$ and 
$a_{11}$, $g_{22}$ which are free.

\item{\underline{Case D3b}}
All the parameters are zero except $k_{222}=1,\ h_{22}=2,\ f_{22}=2$ 
and  $g_{22}, a_{11}, d^a_{21}, \C{d}^a_{21}$ satisfy
$$
4 g_{22} + N^2 a_{11} d^a_{21} \C{d}^a_{21} = 0\ .
$$
\end{description}                                 

\section{}\setcounter{equation}{0}   
Imposing $a_2=0 ,\  b_{21}=1,\ \C{b}_{21}=0$, we obtain
the chiral superalgebra
\bea
\{ \n^i_{\a}, \n^j_{\da} \}&=& a_{11}\q_{[11]}^{ij} \n_{\a\da}
                 \la{Ea} \\[5pt]
\{ \n^i_{\a}, \n^j_{\b} \}&=& 
     N a_{11}c_{11} \q_{[11]}^{ij}M_{\a\b}
   +\e_{\a\b}\left( \left(t_{[21]}^{ikj}-t_{[21]}^{jki}\right)Y^{k}
         +b_{22} t_{[22]}^{ikjl} Z^{kl} \right)
                   \la{Eb} \\[5pt]
\{ \n^i_{\da}, \n^j_{\db} \}&=& \e_{\da\db}\C{b}_{22} t_{[22]}^{ikjl} Z^{kl}
                        \la{Ebb} \\[5pt]
\left[ \n^i_{\a}, \n_{\b\db}  \right] &=&
     \e_{\a\b}\left(c_2 t_{[2]}^{ij}+c_{11}\q_{[11]}^{ij}\right)\n^j_{\db}
                     \la{Ec} \\[5pt]
\left[ \n^i_{\da}, \n_{\b\db} \right]&=&0    \la{Ebc} \\[5pt]
\left[ \n_{\a\da}, \n_{\b\db} \right]&=&0     \la{Er}  \\[10pt]
\left[ Y^i,\n_{\a\da} \right]&=&0
                         \\[5pt]             
\left[ Z^{ij},\n_{\a\da} \right]&=&0   \la{EZD}\\[10pt]   
\left[ Y^i,\n^j_{\a} \right]&=& 
          N a_{11}c_{11} t_{[21]}^{ikj} \n^k_{\a}
                       \la{EYS} \\[5pt]              
\left[ Y^i,\n^j_{\da} \right]&=& a_{11} c_2 t_{[21]}^{ikj} \n^k_{\da}
                          \\[5pt]              
\left[ Z^{ij},\n^k_{\a} \right]&=& h_{22} t_{[22]}^{jkil} \n^l_{\a}
                           \\[5pt]              
\left[ Z^{ij},\n^k_{\da} \right] &=& \C{f}_{22} t_{[22]}^{jkil} \n^l_{\da}
                           \\[5pt]              
[ Y^i, Y^j ]&=&  - N a_{11} c_{11} b_{22} t_{[22]}^{imjn}Z^{mn}
                          \\[5pt]              
[ Z^{ij},Y^k ]&=& h_{22} t_{[22]}^{jkil} Y^l
                         \\[5pt]                
[ Z^{ij},Z^{kl} ]&=& k_{222} t_{[222]}^{jkmiln}Z^{mn}\ \ .
\eea
Here $a_{11}$ and $c_{11}$ are free and the remaining parameters satisfy
the constraints
\bea
&& b_{22} \C{f}_{22} =  b_{22} h_{22} =  b_{22} k_{222} = 0
\\[5pt]
&& \C{b}_{22}  c_2 = \C{b}_{22} \C{f}_{22} =  \C{b}_{22} h_{22} 
     =  \C{b}_{22} k_{222} = 0
\\[5pt]
&& h_{22}(h_{22}-2k_{222}) = \C{f}_{22}(\C{f}_{22}-2k_{222}) 
    = c_2(\C{f}_{22}- h_{22}) = 0\ .
\eea
We find six essentially different subcases 
\begin{description}                                 

\item{\underline{Case E1a}}
The parameters $h_{22}=k_{222}=\C{f}_{22}=c_2=0$, 
and $a_{11}$,  $b_{22}$, $\C{b}_{22}$, $c_{11}$ are free.

\item{\underline{Case E1b}}
The parameters $h_{22}=k_{222}=\C{f}_{22}=\C{b}_{22}=0$, 
and $a_{11}$,  $b_{22}$, $c_2$, $c_{11}$ are free.

\item{\underline{Case E2a}}
Here $k_{222}=1$, $a_{11}$,  $c_{11}$ and $c_2$
are free and the remaining parameters are zero.

\item{\underline{Case E2b}}
Here $k_{222}=1\ ,\ \C{f}_{22}=2$, $a_{11}$, $c_{11}$ are free
and the remaining parameters are zero.

\item{\underline{Case E3a}}
Here $k_{222}=1,\ h_{22}=2$, 
$a_{11}$,  $c_{11}$ are free and the remaining parameters are zero.

\item{\underline{Case E3b}}
Here $k_{222}=1,\ h_{22}=2, \C{f}_{22}=2$,
$a_{11}$, $c_{11}$, $c_2$ are free and the remaining parameters are zero.
 
\end{description}                                  

\section{}\setcounter{equation}{0}      

Class F has the following basic relations, 
$a_2=0 ,\  b_{21}=0 ,\ \C{b}_{21}=0$,  which yield the superalgebra
\bea
\{ \n^i_{\a}, \n^j_{\da} \}&=& a_{11}\q_{[11]}^{ij} \n_{\a\da}
    \la{Fa} \\[5pt]
\{ \n^i_{\a}, \n^j_{\b} \}&=& \e_{\a\b}b_{22} t_{[22]}^{ikjl} Z^{kl}
      \la{Fb} \\[5pt]
\{ \n^i_{\da}, \n^j_{\db} \}&=& \e_{\da\db}\C{b}_{22} t_{[22]}^{ikjl} Z^{kl}
      \la{Fbb} \\[5pt]
\left[ \n^i_{\a}, \n_{\b\db}  \right] &=&
     \e_{\a\b}\left(c_2 t_{[2]}^{ij}+c_{11}\q_{[11]}^{ij}\right)\n^j_{\db}
    \la{Fc} \\[5pt]
\left[ \n^i_{\da}, \n_{\b\db} \right]&=&
\e_{\da\db}\left(\C{c}_2 t_{[2]}^{ij}+\C{c}_{11}\q_{[11]}^{ij}\right)\n^j_{\b}
    \la{Fbc} \\[5pt]
\left[ \n_{\a\da}, \n_{\b\db} \right]&=&
  c_2 \C{c}_2 \left( \e_{\a\b}M_{\da\db}+ \e_{\da\db}M_{\a\b}\right) 
    \la{Fr}  \\[10pt]
\left[ Y^i,\n_{\a\da} \right]&=&0
        \\[5pt]             
\left[ Z^{ij},\n_{\a\da} \right]&=&0 \la{FZD}\\[5pt]   
\left[ Y^i,\n^j_{\a} \right]&=&
              \left(d_3 t_{[3]}^{ijk}
                   +d_{21}^a t_{[21]}^{ijk}
                   +d_{21}^b t_{[21]}^{ikj}\right)\n^k_{\a}
      \la{FYS} \\[5pt]              
\left[ Y^i,\n^j_{\da} \right]&=&
              \left(\C{d}_3 t_{[3]}^{ijk}
                   +\C{d}_{21}^a t_{[21]}^{ijk}
                   +\C{d}_{21}^b t_{[21]}^{ikj}\right)\n^k_{\da}
        \\[5pt]              
\left[ Z^{ij},\n^k_{\a} \right]&=&
       f_{22} t_{[22]}^{jkil} \n^l_{\a}
        \\[5pt]              
\left[ Z^{ij},\n^k_{\da} \right] &=& \C{f}_{22} t_{[22]}^{jkil} \n^l_{\da}
        \\[5pt]              
[ Y^i, Y^j ]&=&  g_{22} t_{[22]}^{imjn}Z^{mn}
        \\[5pt]              
[ Z^{ij},Y^k ]&=& h_{22} t_{[22]}^{jkil} Y^l
       \\[5pt]                
[ Z^{ij},Z^{kl} ]&=& k_{222} t_{[222]}^{jkmiln}Z^{mn}\ \ .
\eea
For $k_{222}=h_{22}=0$, i.e. case F1, we will limit ourselves to giving 
the conditions which have to be fulfilled.
In the cases F2, F3 where  $k_{222}=1$, 
we give a more precise discussion. 
There are many subcases which we have classified as follows
\bea
{\rm{subcase\ a\ :}}&& a_{11}\neq 0
        \nonumber\\
{\rm{subcase\ b\ :}}&& a_{11}=0,\  f_{22} =0,\ \C{f}_{22}=0
        \nonumber\\
{\rm{subcase\ c\ :}}&& a_{11}=0,\  f_{22} =2,\ \C{f}_{22}=0
        \nonumber\\
{\rm{subcase\ d\ :}}&& a_{11}=0,\  f_{22} =2,\ \C{f}_{22}=2 \ \ .
        \la{subsubcases}
\eea
With this in mind, we find nine essentially different subcases.

\begin{description}   

\item{\underline{Case F1}}
With $h_{22}=k_{222}=0$,  
which implies $f_{22}=\C{f}_{22}=0$,
the twenty conditions to be fulfilled are
\bea 
b_{22} d_{21}^b=0  
&,& \ \ b_{22} d_{3}=0
       \nonumber\\
\C{b}_{22} \C{d}_{21}^b=0 
&,& \ \ \C{b}_{22} \C{d}_{3}=0
       \nonumber\\
a_{11} c_{2}=0  
&,& \ \ a_{11} c_{11}=0
       \nonumber\\
a_{11}\C{c}_2=0 
&,& \ \ a_{11} \C{c}_{11}=0
       \nonumber\\
a_{11} d_{21}^a  =0  
&,& \ \   a_{11} \C{d}_{21}^a =0
       \nonumber\\
c_2\C{c}_{2}-N^2 c_{11}\C{c}_{11}=0 
&,& \ \ \C{b}_{22} c_{2} - b_{22} \C{c}_{2}=0
       \nonumber\\
c_{2} (\C{d}_{3} - d_{3})=0 
&,&\ \ \C{c}_{2} (\C{d}_{3} - d_{3})=0
       \nonumber\\
c_{2} \C{d}_{21}^a - N c_{11} d_{21}^a   =0  
&,& \ \ 
\C{c}_{2} \C{d}_{21}^b - N \C{c}_{11} d_{21}^b =0 
       \nonumber\\
c_{2} d_{21}^b - N c_{11} \C{d}_{21}^b =0  
&,& \ \ 
 \C{c}_{2} d_{21}^a - N \C{c}_{11} \C{d}_{21}^a  =0
       \nonumber \\
d_3^2 - N^2 d_{21}^a d_{21}^b =0 
&,&\ \ 
\C{d}_3^2 - N^2 \C{d}_{21}^a \C{d}_{21}^b  =0\ .
\la{caseF1}
\eea
This leads to a rather long, easy but uninteresting 
discussion which we will not give.

\item{\underline{Case F2a}}
Here $k_{222}=1$, $a_{11}\neq 0$ ,
$d_{21}^b$, $\C{d}_{21}^b$, $f_{22}$ and $\C{f}_{22}$ satisfy 
the following conditions .
\bea
f_{22}(f_{22}-2)&=&0             \nonumber\\
\C{f}_{22}(\C{f}_{22}-2)&=&0     \nonumber\\
f_{22}d_{21}^b&=&0               \nonumber\\
\C{f}_{22}\C{d}_{21}^b&=&0               
\label{casef2a}
\eea
and the remaining parameters are zero.

\item{\underline{Case F2b}}
Here $k_{222}=1$, 
$c_2$, $c_{11}$, $\C{c}_2$, $\C{c}_{11}$,
$d_{21}^a$,  $d_{21}^b$, $d_{3}$, $\C{d}_{21}^a$,  
$\C{d}_{21}^b$, $\C{d}_{3}$ satisfy the conditions
\bea
c_2\C{c}_{2}-N^2 c_{11}\C{c}_{11}=0 
&,&
       \nonumber\\
c_{2} (\C{d}_{3} - d_{3})=0 
&,&\ \ \C{c}_{2} (\C{d}_{3} - d_{3})=0
       \nonumber\\
c_{2} \C{d}_{21}^a - N c_{11} d_{21}^a   =0  
&,& \ \ 
\C{c}_{2} \C{d}_{21}^b - N \C{c}_{11} d_{21}^b =0 
       \nonumber\\
c_{2} d_{21}^b - N c_{11} \C{d}_{21}^b =0  
&,& \ \ 
 \C{c}_{2} d_{21}^a - N \C{c}_{11} \C{d}_{21}^a  =0
       \nonumber \\
d_3^2 - N^2 d_{21}^a d_{21}^b =0 
&,&\ \ 
\C{d}_3^2 - N^2 \C{d}_{21}^a \C{d}_{21}^b  =0
\la{condF2b}
\eea
and the remaining parameters are zero.

\item{\underline{Case F2c}}
All the parameters are zero except $k_{222}=1$, $f_{22}=2$ and
$c_{11}$,  $\C{c}_{11}$, $\C{d}_3$,
$\C{d}_{21}^a$ and $\C{d}_{21}^b$ which
satisfy
\bea
c_{11}\C{c}_{11}=0
&,&\ \ 
  \C{d}_3^2 - N^2 \C{d}_{21}^a \C{d}_{21}^b  =0
       \nonumber\\
c_{11}\C{d}_{21}^b=0
&,&\ \ 
   \C{c}_{11}\C{d}_{21}^a=0
\ \ .
\la{condF2c}
\eea

\item{\underline{Case F2d}}
All the parameters are zero except $k_{222}=1$, 
$f_{22}=\C{f}_{22}=2$ and $c_2$,
$c_{11}$, $\C{c}_{2}$, $\C{c}_{11}$ 
which satisfy the condition
\be 
c_{2}\C{c}_{2}-N^2 c_{11}\C{c}_{11}=0\ .
\la{condF2d}
\ee

\item{\underline{Case F3a}}
All the parameters are zero except $k_{222}=1$, $h_{22}=2$, 
$a_{11}\neq 0$ and
$d_{21}^b$, $\C{d}_{21}^b$, $f_{22}$, $\C{f}_{22}$
and $g_{22}$ which satisfy the conditions
\bea
f_{22} (f_{22} - 2)=0
&,&\ \ 
  \C{f}_{22} (\C{f}_{22} - 2)=0
      \nonumber\\
d_{21}^b (f_{22} - 2)=0
&,&\ \ 
  \C{d}_{21}^b ( \C{f}_{22} - 2)=0
      \nonumber\\
f_{22} g_{22}=0
&,&\ \ 
  \C{f}_{22} g_{22}=0 \ \ .
\la{caseF3a }
\eea

\item{\underline{Case F3b}}
All the parameters are zero except for $k_{222}=1$, $h_{22}=2$, $g_{22}$ 
and $c_2$, $c_{11}$, $\C{c}_2$, 
and $\C{c}_{11}$ which satisfy the condition
\be
c_{2} \C{c}_{2}  - N^2 c_{11} \C{c}_{11}=0\ \ .
\la{caseF3b}
\ee

\item{\underline{Case F3c}}
All the parameters are zero except  $k_{222}=1$, $h_{22}=2$, $f_{22}=2$
and $c_{11}$,  $\C{c}_{11}$,  $d_{21}^a$ and
$d_{21}^b$ which satisfy the conditions
\be
c_{11} \C{c}_{11} =0
        \quad ,\quad
c_{11} d_{21}^a =0
        \quad ,\quad 
\C{c}_{11} d_{21}^b  =0
        \la{caseF3ccond}
\ee
and the dependent parameter
\be
g_{22} =\fr{N}{4} d_{21}^a d_{21}^b \ \ .
    \la{caseF3c}
\ee

\item{\underline{Case F3d}}
All the parameters are zero except for $k_{222}=1$, $h_{22}=2$, 
$f_{22}=\C{f}_{22}=2$ and $c_{11}$,
$c_{2}$,  $\C{c}_{11}$, $\C{c}_{2}$, 
$d_{21}^a$,  $d_{21}^b$ $\C{d}_{21}^a$ and
$\C{d}_{21}^b$ which satisfy the conditions
\bea
c_{2} \C{d}_{21}^a - N c_{11} d_{21}^a   =0  
&,& \ \ 
\C{c}_{2} \C{d}_{21}^b - N \C{c}_{11} d_{21}^b =0 
       \nonumber\\
c_{2} d_{21}^b - N c_{11} \C{d}_{21}^b =0  
&,& \ \ 
 \C{c}_{2} d_{21}^a - N \C{c}_{11} \C{d}_{21}^a  =0
       \nonumber \\
c_2\C{c}_{2}-N^2 c_{11}\C{c}_{11}=0 
&,& \ \ 
  d_{21}^a d_{21}^b -  \C{d}_{21}^a \C{d}_{21}^b=0
\la{caseF3dcond}
\eea
and the dependent parameter
\be
g_{22} =\fr{N}{4} d_{21}^a d_{21}^b \ \ .
\la{caseFd}
\ee
\end{description}   

\goodbreak
\begin{thebibliography}{99}

\bibitem{HLS} R. Haag, J.T. Lopuszansky, M. Sohnius,
{\it All possible generators of supersymmetries of the S-matrix}, 
{\sl Nucl. Phys.\/ \bf B88} (1975) 257-274
%
\bibitem{N} W. Nahm,
{\it Supersymmetries and their representations}, 
{\sl Nucl. Phys. \/ \bf B135} (1978) 149-166
%
\bibitem{H} S. Hewson,
{\it{Generalised supersymmetry and $p$-brane actions}},
{\sl Nucl. Phys. \/ \bf B501} (1997) 445-468, hep-th/9701011 
%
\bibitem{DO} C. Devchand and V. Ogievetsky,
{\it Conserved currents for unconventional supersymmetric couplings of 
self-dual gauge fields},
{\sl Phys. Lett.} {\bf B367} (1996) 140-144, hep-th/9510235;
C. Devchand and V. Ogievetsky,
{\it Interacting fields of arbitrary spin and $N{>}4$ 
supersymmetric self-dual Yang--Mills equations},
{\sl Nucl. Phys.} {\bf B481} (1996) 188-214, hep-th/9606027
%
\bibitem{DL} C. Devchand and O. Lechtenfeld,
{\it{Extended self-dual Yang Mills from $N=2$ string}},  
{\sl Nucl. Phys. \/ \bf B516} (1998) 255-272, hep-th/9712043
%
\bibitem{AC} D.V. Alekseevsky and V. Cort\'es,
{\it  Classification of $N$-(super)-extended Poincar\'e algebras 
and bilinear invariants of the spinor representation of $Spin(p,q)$},  
{\sl Commun. Math. Phys.} {\bf 183} (1997) 477-510
%
\bibitem{FV} E.S. Fradkin and M.V. Vasiliev,
{\it{Candidate for the role of higher spin symmetry}},
{\sl Ann. Phys.} {\bf 177} (1987) 63-112;
M.V. Vasiliev,
{\it{Consistent equations for interacting massless fields of all spins
in the first order in curvatures}},
{\sl Ann. Phys.} {\bf 190} (1989) 59-106
%
\bibitem{DN} C. Devchand and J. Nuyts,
{\it{Selfduality in generalized Lorentz superspaces}},
{\sl Phys. Lett.} {\bf B404} (1997) 259-263, hep-th/9612176;
%
C. Devchand and J. Nuyts,
{\it{Supersymmetric Lorentz covariant hyperspaces and 
self-duality equations in dimensions greater then (4/4)}},
{\sl Nucl. Phys. \/ \bf B503} (1997) 627-656, hep-th/9704036;
%
C. Devchand and J. Nuyts,
{\it{Lorentz covariant spin two superspaces}},
{\sl Nucl. Phys. \/ \bf B527} (1998) 479-498, hep-th/9804052
%
\bibitem{CM} S. Coleman and J. Mandula, 
{\it{All possible symmetries of the S-matrix}},  
{\sl Phys. Rev. \/ \bf159} (1967) 1251-1256
%
\end{thebibliography}
\end{document}